\documentclass[namedreferences]{SolarPhysics}
\usepackage[optionalrh]{spr-sola-addons} 

\usepackage{graphicx}        
\usepackage{natbib}          
\usepackage{color}           
\usepackage{url}             

\begin{document}

\begin{article}

\begin{opening}

\title{Very High Resolution Solar X-ray Imaging Using Diffractive Optics}

\author{B.~R.~\surname{Dennis}$^{1}$\sep
        G.~K.~\surname{Skinner}$^{2,3}$\sep
        M.~J.~\surname{Li}$^{4}$\sep
        A.~Y.~\surname{Shih}$^{1}$
       }
\runningauthor{Dennis et al.}
\runningtitle{Diffractive X-ray optics}

   \institute{$^{1}$ Code 671, NASA Goddard Space Flight Center
                     email: \url{brian.r.dennis@nasa.gov} \\
              $^{2}$ Code 661
                     email: \url{gerald.k.skinner@nasa.gov} \\
             $^{3}$ CRESST and Univ of Maryland, College Park\\
              $^{4}$ Code 553
                     email: \url{mary.j.li@nasa.gov} \\
             }

\begin{abstract}
This paper describes the development of X-ray diffractive optics for imaging solar flares with better than 0.1 arcsec angular resolution. X-ray images with this resolution of the $\ge10$ MK plasma in solar active regions and solar flares would allow the cross-sectional area of magnetic loops to be resolved and the coronal flare energy release region itself to be probed. The objective of this work is to obtain X-ray images in the iron-line complex at 6.7 keV observed during solar flares with an angular resolution as fine as 0.1 arcsec - over an order of magnitude finer than is now possible. This line emission is from highly ionized iron atoms, primarily Fe {\sc xxv}, in the hottest flare plasma at temperatures in excess of $\approx$10 MK. It provides information on the flare morphology, the iron abundance, and the distribution of the hot plasma. Studying how this plasma is heated to such high temperatures in such short times during solar flares is of critical importance in understanding these powerful transient events, one of the major objectives of solar physics. We describe the design, fabrication, and testing of phase zone plate X-ray lenses with focal lengths of $\approx$100 m at these energies that would be capable of achieving these objectives. We show how such lenses could be included on a two-spacecraft formation-flying mission with the lenses on the spacecraft closest to the Sun and an X-ray imaging array on the second spacecraft in the focal plane $\approx$100 m away. High resolution X-ray images could be obtained when the two spacecraft are aligned with the region of interest on the Sun. Requirements and constraints for the control of the two spacecraft are discussed together with the overall feasibility of such a formation-flying mission.

\end{abstract}
\end{opening}

\section{Introduction}
     \label{Sect:Introduction}

High resolution X-ray imaging has been discussed extensively in astrophysics over many years.  One of the most exciting prospects with the 0.1 to 1 micro-arcsecond resolution that should be possible with diffractive optics is to be able to probe the space-time at the event horizon of super-massive black holes, \textit{e.g.} \cite{2001A&A...375..691S, 2009SPIE.7437E..17S}. However, achieving this resolution requires lenses with focal lengths of hundreds or thousands of km. Their use would require flying X-ray detectors on one spacecraft and lenses on a second spacecraft separated from the first by the focal length in the direction of the source of interest. Formation flying with such large spacecraft separations and the stabilization and alignment knowledge necessary to generate such high-resolution images would be a major advance on existing capabilities.

We show here that a less demanding form of the same technology can contribute in heliophysics to the frontier area of solar flare research. The high resolution X-ray imaging of the Sun possible with diffractive optics on a scale more modest than that considered for astrophysical applications offers the possibility of obtaining significantly finer angular resolution than is possible with the conventional diffraction-limited reflective optics used at longer wavelengths. Even a 1-m diameter mirror is diffraction limited at optical wavelengths at $\approx0.1$ arcsec. The Advanced Technology Solar Telescope (ATST) in Hawaii will operate at the diffraction limit of a 4-m diameter mirror in the near infra-red wavelength range of 900 - 2500 nm to achieve 0.03 arcsec resolution at 500 nm and 0.08 arcsec at 1.6 $\mu$m over a field of view of 2 to 3 arcmin \citep{2011ASPC..437..319K}. Because of the much shorter wavelengths of X-rays (1 \AA ~= 0.1 nm), diffraction-limited X-ray optics of only 1 cm in diameter could equal or even improve on this angular resolution.

 High resolution imaging of solar X-rays would be invaluable in both soft X-rays ($\lesssim$10 keV) from hot plasma and hard X-rays ($\gtrsim$ 10 keV)) from nonthermal distributions of energetic electrons. Soft X-ray observations with 0.1 arcsec resolution of the $\gtrapprox$10 MK plasma in solar active regions and solar flares would allow the cross-sectional area of magnetic loops to be resolved and could allow the coronal flare energy release region itself to be probed on physically meaningful spatial scales. At higher energies, high resolution imaging would be possible of hard X-rays produced by beams of nonthermal electrons as they stream down from the acceleration site in the corona into the higher density regions of the lower corona and chromosphere. The structure of the compact footpoint X-ray sources at the 0.1 arcsec level will provide information on the details of the coronal acceleration process itself. The ability to detect the relatively weak coronal emission from the electron beams in the presence of the bright chromospheric footpoint sources (dynamic range requirement of $\gtrapprox$100:1) will also allow the acceleration site to be studied in unprecedented detail.

\begin{figure}
   \begin{center}
   \begin{tabular}{c}
\includegraphics[  angle=0, trim = 0mm 0mm 0mm 0mm, clip, width=11cm]{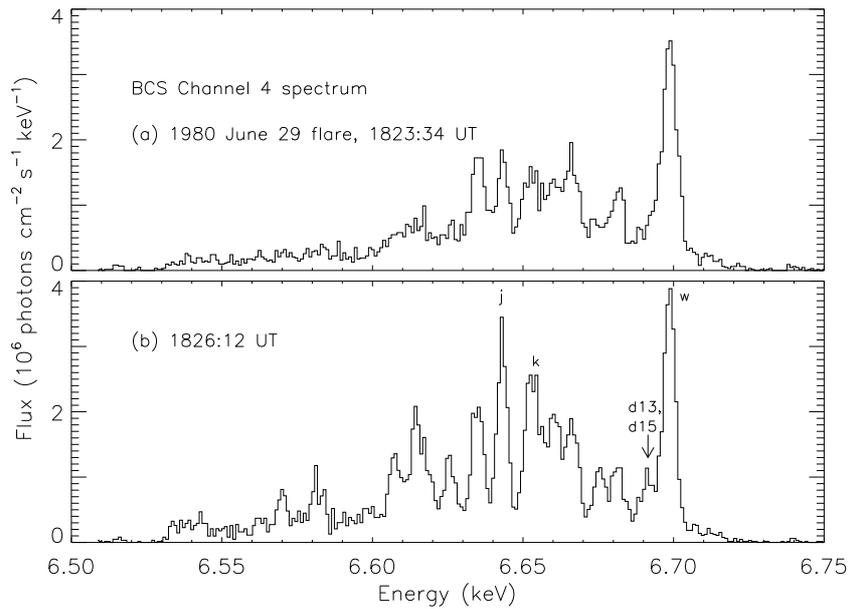}
    \end{tabular}
    \end{center}
   \caption[example]
   {High resolution X-ray spectra of a solar flare showing the complex of lines between 6.5 and 6.75 keV. The lines are from highly ionized iron atoms in a plasma at a temperature of $\approx$20 MK, with the most prominent line from Fe {\sc xxv} at 6.699 keV that is labeled with a 'w' in the lower plot using the naming convention of \cite{1972MNRAS.160...99G}. The spectrum was recorded with the Bent Crystal Spectrometer (BCS) on the Solar Maximum Mission during a flare in 1980 June 29. }
   \label{fig:BCS_spectra}
   \end{figure}

 Although there is great interest in X-ray imaging over a broad energy range extending into the nonthermal domain above 10 keV, technical issues limit the technique using diffractive X-ray optics to a very narrow energy range. Consequently, for this initial demonstration of high resolution solar X-ray imaging we have chosen to concentrate on the soft X-ray range, specifically
the  group of Fe emission lines between 6.6 and 6.7 keV. The strongest line is typically
the `w'  Fe {\sc xxv} resonance
line at 6.699 keV (1.851\AA) as seen in  Figure \ref{fig:BCS_spectra}. By using detectors with relatively modest energy resolution or by absorbing lower energy photons, it is possible to arrange that the diffractive optics image of a flare is dominated by emission in this line complex. In this way we are able to make images almost free from blurring due to photons at other energies that come to a focus at different distances due to the chromatic aberration intrinsic to diffractive optics. We propose to use this line complex to facilitate the first demonstration of high resolution imaging using diffractive X-ray optics and readily available detectors. It will be possible to extend this technique to higher energies in the future but because the solar flare X-ray emission above 10 keV is all bremsstrahlung continuum with no narrow lines, detectors with high energy resolution will be required to isolate the energies in the narrow range that come to a focus at the detector.

\subsection{Solar Flare X-ray Emission}
     \label{Sect:Science}

X-rays from solar flares in the energy range between $\approx$1 and 10 keV are emitted from hot plasma with temperatures from $\approx$$10^6$ K (1 MK) to sometimes in excess of 50 MK. Detailed studies of this radiation, both spectroscopically and by imaging, provide crucial information on the nature of the heated plasma, and can provide clues to help us understand how such a large volume of plasma can be heated to such high temperatures so rapidly during a flare.

The solar spectrum in this energy range is a combination of both line and continuum emission. The many narrow lines are emitted by transitions of atoms of the different elements of the plasma in the solar atmosphere in various stages of ionization; the continuum emission is from both free-free and free-bound interactions of electrons with atomic nuclei that produce bremsstrahlung and recombination radiation, respectively. The most detailed catalog of lines in the 3-10 keV range is given by \cite{2004ApJ...605..921P, 2008A&A...490..823P}. Both line and continuum spectra can be synthesized using the CHIANTI database and software \citep{1997A&AS..125..149D, 2009A&A...498..915D} for a range of possible temperatures, plasma abundances, and excitation and ionization conditions. The line of greatest interest here is the Fe {\sc xxv} resonance line at 6.699 keV (1.851 \AA). It is the strongest and most prominent line of the complex of iron lines between 6.5 and 6.7 keV (Figure \ref{fig:BCS_spectra}).  This line was named the `w' line by \cite{1972MNRAS.160...99G} and arises from the $1s^{2~1}S_0 - 1s2p ^{1}P_{1}$ transition of helium-like iron nuclei, i.e., with just two electrons remaining of the original 26 of an unionized iron atom. It is emitted from plasma at temperatures above $\approx$10 MK, and hence provides information on the hottest plasma generated in a solar flare. During the flare impulsive phase, the line shows a broadening that is commonly attributed to turbulence in the plasma resulting from the impact
of nonthermal electrons. The broadening may amount to 0.03 \AA ~(0.01 keV or 10 eV), equivalent
to a few hundred km s$^{-1}$. Nevertheless, the line is always sufficiently narrow to allow imaging with better than 0.1 arcsec resolution using diffractive optics.

It is clear from images taken with existing X-ray instruments and those in different wavelength bands that there are features that are unresolved at the currently resolvable angular scale of $\approx$1 arcsec. The ubiquitous magnetic loops observed before, during, and after flares appear to be significantly narrower than 1 arcsec. The magnetic reconnection process that is believed to lead to the energy release from the coronal magnetic field that heats the plasma takes place on scales much smaller than 1 arcsec. The total energy and density of the thermal plasma can be estimated from the emission measure and source volume as revealed by the X-ray images but the values so obtained are subject to a large, order-of-magnitude, uncertainty because of the unknown ``filling factor,'' the ratio of the apparent volume of the plasma to the actual volume. Only by making images with higher angular resolution can this uncertainty be diminished and ultimately eliminated once the finest plasma structures are resolved. These unknowns then set the goal of the current effort to image in soft X-rays with better than 0.1 arcsec resolution, fully an order of magnitude better than what has been achieved to date.

\subsection{Current Capabilities and Limitations}

It is clear that the finest possible angular resolution is required to fully understand the heating of the flare plasma. To date, the highest angular resolution for solar observations in this energy range is  $\approx$1 arcsec (full width at half maximum, FWHM) achieved with the X-ray Telescope (XRT) on Hinode \citep{2007SoPh..243...63G, 2007PASJ...59S.853W} using a single monolithic grazing-incidence mirror with two reflections.  With grazing incidence optics, angular resolution is limited by surface figure and, where nested optics are employed, by alignment considerations.  As pointed out by \cite{2011SPIE.8148E..22D}, for example, conventional high energy optics techniques cannot obtain the required $<$0.1 arcsec resolution.


The \textit{Ramaty High Energy Solar Spectroscopic Imager} (RHESSI) has achieved $\approx$2 arcsec angular resolution using Fourier-transform imaging at energies extending from $\approx$3 keV up to 100 keV, and with progressively poorer resolution up to 17 MeV \citep{2002SoPh..210....3L}. Bi-grid rotating modulation collimators are used with 1.5 m separation between grids and a finest slit pitch of 34 $\mu$m. Higher resolution would be possible with finer slits and/or greater grid separation  but at the expense of field of view. In addition, the achievable dynamic range in any one image using this technique is limited to less than $\approx$50:1 by side lobes in the response function. This makes it difficult, for example, to image the relatively weak and extended coronal X-ray sources in the presence of intense compact footpoint sources.

\section{Diffractive X-ray Optics}
     \label{Sect:Diff_Opt}
The angular resolution possible with reflecting optics is critically dependent on the accuracy of the polishing of the mirror surfaces and on their alignment. In contrast, the tolerances necessary to manufacture diffractive lenses capable of ultra-high angular resolution are much more relaxed. This is because diffractive focusing depends on modulation of the phase and/or amplitude of radiation by an optical element working in transmission at normal incidence. This means that the lenses can be relatively thin, and alignment precision is not a serious issue.

\cite{2011SPIE.8148E..22D} has proposed the use of a photon sieve, a form of Fresnel zone plate, to achieve high angular resolution in the extreme ultraviolet. We concentrate here on Fresnel lenses that can offer higher efficiency than zone plate variants and suffer less from diffraction into spurious orders. The principles of phase Fresnel lenses and their development for X-ray and gamma-ray astronomy have been discussed in a series  of papers   \citep{2001A&A...375..691S, krizmanic05a, Krizmanic07,  2009SPIE.7437E..17S}.

\subsection{Lens Design Considerations}
\label{sect:design}

The focussing ability of diffractive optics requires that radiation that passes through different parts of the lens must all arrive at the focal point with identical phase. This can be achieved if the thickness, $t$, of the lens with a focal length, $f$, is the following function of the radial distance, $r$, from its center:


	\begin{eqnarray}
t(r) & = & \left( r^2 / (2f\delta) \pmod {\lambda} \right) + t_0\\
& = & \left( r^2 / (2f\lambda) \pmod {1} \right) t_{2\pi} +t_0,
    \label{eqn:t2pi}
   \end{eqnarray}

\noindent where $\lambda$ is the wavelength and the refractive index is $\mu = 1 - \delta$ \citep[see][]{Krizmanic07}. The addition of the $t_0$ term
reflects the fact that, apart from absorption, the presence of a constant thickness substrate does not affect the focussing properties, just the throughput. In Equation \ref{eqn:t2pi}, we have written $t_{2\pi}$ for the thickness of material, $\lambda/\delta$,  that changes the phase by $2\pi$. Conveniently, in the soft X-ray band, this quantity is typically a few $\mu$m leading to structures well matched to fabrication by Micro-Electro-Mechanical Systems (MEMS) engineering technology.

\begin{figure}
   \begin{center}
   \begin{tabular}{c}
 \includegraphics[  angle=-90, trim = 5mm 10mm 20mm 10mm, clip, width=11cm]{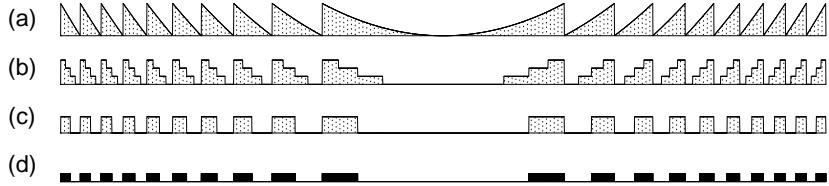}\\
   \end{tabular}
   \end{center}
   \caption[example]
   {   (a) The ideal cross-section through a Phase Fresnel Lens (PFL) with a maximum thickness (t) of
   $t_0$ (which may be zero) $ +  \hspace{0.05in} t_{2\pi}$. For X-rays, the refractive index is less than one
   ($\delta$ is positive) so the configuration shown corresponds to a converging lens. (b) A four-level approximation to a PFL profile with a thickness of $t_0 + 0.75 \hspace{0.05in} t_{2\pi}$. (c) A two-level approximation (phase zone plate, PZP) with a thickness of $t_0 + 0.5 \hspace{0.05in} t_{2\pi}$. (d) A zone plate, in which alternate zones are opaque to the radiation. }
   \label{fig:profiles}
   \end{figure}

The ideal profile corresponding to Equation \ref{eqn:t2pi}, as illustrated in Figure \ref{fig:profiles}a, produces an optic known as a phase Fresnel lens (PFL). It can be  replaced by a multi-step approximation (Figure \ref{fig:profiles}b), the limiting case being where there are just two thicknesses - $t_0$
 (which may be zero) and $t_0 + t_{2\pi}/2$ - as illustrated in Figure \ref{fig:profiles}c. In a further approximation, it is possible simply to block the radiation for which the phase is not within $\pm \pi/2$ of the ideal, leading to the zone plate structure shown in Figure \ref{fig:profiles}d. For that reason, a two-level diffractive lens is often referred to as a phase zone plate or PZP. The ideal efficiencies, defined as the flux in the first order focus as a fraction of the total incident flux, are listed in Table \ref{table:eff} for the different cases.

\begin{table}[htdp]
\caption{The ideal efficiency of diffraction into the first-order focus for lenses with profiles such as those illustrated in Figure \ref{fig:profiles}. Absorption is assumed to be negligible. }
\begin{tabular}{lccccc}
\hline
                &  phase     &   &  & phase   &   \\
Approximation & Fresnel lens & $n$-levels & 4-levels  & zone plate & zone plate \\
 & (PFL)        &  &  & (PZP)      & (ZP)\\

Figure \ref{fig:profiles}  & (a)  &  &     (b) & (c) &  (d) \\
\hline
Efficiency   & 1 &  $\left[\frac{n}{\pi}\sin(\frac{\pi}{n})\right]^2$ & 0.811 & 0.405 & 0.101 \\               \hline
\end{tabular}
\label{table:eff}
\end{table}

Irrespective of the profile adopted, at the edge of the lens the structure is periodic with the minimum period given by
\begin{equation}
p_{\mathrm{min}} =  2f\lambda/d,  \label{eqn:pmin}
\end{equation}
where $d$ is the lens diameter. The focal length $f$ can be written  in terms of the photon energy, $E$, in the range that is of particular interest here as:
\begin{equation}
f   =  102  \left(\frac{p_{\mathrm{min}}}{1.9\mbox{ $\mu$m}}\right)  \left(\frac{E}{6.65\mbox{ keV}}\right)  \left( \frac{d}{20\mbox{  mm}}\right) \mbox{ m} \label{eqn:f}.
\end{equation}
For a diffraction-limited lens, the angular diameter of the Airy disk is :
 \begin{equation}
            \theta_{\mathrm{d}}   =  2.44 \: \lambda /d  \: =  1.22  \: p_{\mathrm{min}}/f.
\end{equation}
For an ideal lens, the Airy disk contains 84\% of the energy in the first order focus. With typical parameters, its angular size is much smaller than the resolution targeted here:
 \begin{equation}
\theta_\mathrm{d}   =  4.7  \left(\frac{E}{6.65\mbox{ keV}}\right)^{-1}  \left( \frac{d}{20\mbox{ mm}}\right)^{-1} \:{\mbox {milli-arcsec}}
\label{eqn:diff}.
 \end{equation}


For our purposes, a more important consideration than the diffraction limit is blurring due to chromatic aberration. Equation \ref{eqn:f} shows that the focal length is proportional to photon energy so in practice the use of such lenses is restricted to observations in which a narrow range of energies is dominant or where the bandwidth is limited in other ways.
If the point spread
function (PSF) of the lens is approximated by the best-fit Gaussian and the spectral line to be imaged is also taken to be Gaussian with a FWHM of $\Delta E$, then the chromatic aberration contribution to the FWHM angular resolution of the lens is found to be
\begin{equation}
   \theta_{\mathrm{c}}  =   5.2 \left( \frac{\Delta E}{E}\right) \;  \left( \frac{d}{20\mbox{ mm}}\right) \;  \left( \frac{f}{100 \mbox{ m}}\right)^{-1} \:{\mbox {arcsec}}.   \label{eqn:chrom}
\end{equation}
Note that the PSF is strongly cusped and the central peak is much sharper than this implies. On the other hand, if the criterion adopted is the width that contains 84\% of the energy (the fraction within the central peak of an Airy disk), then the numerical factor in Equation \ref{eqn:chrom} is more than doubled.

A typical line width due to thermal broadening of the Fe {\sc xxv} 6.699 keV line is about 2 eV, corresponding to a Doppler velocity of $\approx$100~km~s$^{-1}$. There is often additional broadening due to turbulence, and line shifts due to bulk motion, both of which can be several times larger than this. Furthermore, although this spectral line is frequently dominant, it is just one line of many lines in a complex that can be approximated by a Gaussian with a characteristic width of about 100 eV. If the detector is energy resolving and can be used in a photon counting mode, one can attempt to select just those photons that fall within a narrow energy band. However, for Silicon CCDs and similar detectors, the attainable FWHM resolution $\Delta E$ at 6.7 keV is limited to about 150 eV -- wider than the line complex. If we take a width of 100 eV and the reference parameters used in Equations \ref{eqn:f} and \ref{eqn:diff}, the limit due to chromatic aberration is 80 milli-arcsec FWHM.

A further limit to the angular resolution potentially arises because of the spatial resolution of the detector.  For a pixel size $\Delta x$ the corresponding limit is
\begin{equation}
\theta_{\mathrm{s}}  = \frac{\Delta x}{f}  =  21  \left(\frac{\Delta x}{10 {\mbox{ $\mu$m} } }\right)  \left( \frac{f}{100 \mbox{ m}}\right)^{-1} \:{\mbox {milli-arcsec} . \label{eqn:detlim}  }
\end{equation}

The work of \cite{1972JOSA...62..972Y} on aberrations in zone plate imaging can be applied to Fresnel lens optics. It shows that, for systems of interest for solar physics or astrophysics, geometric aberrations are very small, and the only limit on the field of view will be that imposed by the size of a practical detector.

Thus, none of these limits prevent an angular resolution of the order of 100 milli-arcsec from being obtained in imaging observations in the Fe-line complex provided that lenses with $p_{\mathrm{min}}$ of 1.5 -- 2 $\mu$m can be made and used in a configuration with a focal length of $\approx$100 m.


    \begin{figure}
   \centerline{\includegraphics[width=0.5\textwidth]{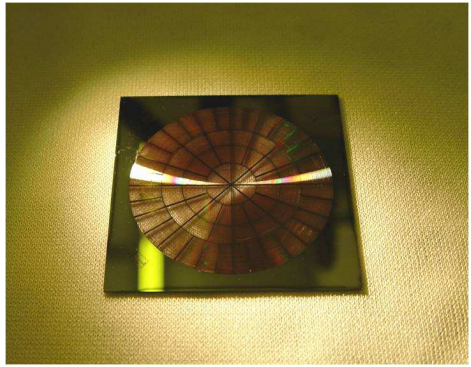}      }
   \caption{Photograph of a 3-cm diameter silicon phase zone plate fabricated in the Detector Development Laboratory (DDL) at Goddard Space Flight Center.}
   \label{fig:PZPphoto}
   \end{figure}

 \begin{figure}
   \centerline{\includegraphics[width=0.5\textwidth]{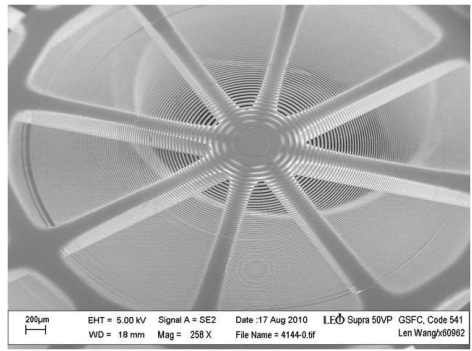}      }
   \caption{\textbf{Scanning electron microscope image of the central area of a demonstration lens showing the circular slits and the radial support ribs.}
                      }
   \label{fig:Lens_SEMimage}
 \end{figure}

\begin{figure}
   \centerline{\includegraphics[width=0.5\textwidth]{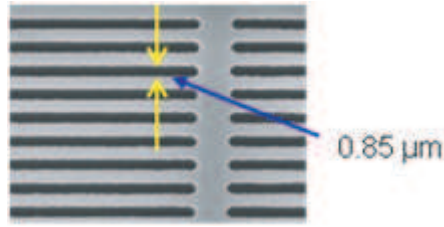}    }
   \caption{\textbf{Scanning electron microscope image of the outer section of a lens showing the finest slits and part of a radial support rib.}
                      }
   \label{fig:SEMslits}
\end{figure}

\subsection{Lenses Fabrication}
\label{sect:fab}

We have initially concentrated on the fabrication of PZP lenses with the cross-section shown in Figure \ref{fig:profiles}c. For a given $t_{2\pi}$, it is easier to obtain a small $p_{\mathrm{min}}$ with this profile than to fabricate the tapered PFL profile (Figure \ref{fig:profiles}a) or a multi-level approximation to one (Figure \ref{fig:profiles}b). For a particular focal length, a smaller $p_{\mathrm{min}}$ allows the lens diameter to be larger (per Equations (\ref{eqn:pmin}) and (\ref{eqn:f})),  and the increased  geometric area can compensate for the lower efficiency. We have fabricated lenses with parameters given in Table \ref{table:params} that are similar to the reference values in  Equation (\ref{eqn:f}) needed to meet the requirements for a flight instrument. A photograph of such a phase zone plate is shown in Figure \ref{fig:PZPphoto} with scanning electron microscope images of the central area in Figure \ref{fig:Lens_SEMimage} and the finest slits at the edge in Figure \ref{fig:SEMslits}.

\begin{table}[htdp]
\caption{Comparison of the parameters of the demonstration lenses with a possible flight design. }
\begin{tabular}{lcccc}
\hline
\multicolumn{2}{c}{Parameter}       & Demonstration              & Possible flight   &     \\
                        &            & laboratory  lenses                    &    design        &      \\
 \hline
Energy            & $E$     &      5.411                        &       6.65     &  keV \\
Focal length   & $f$       &      110.4                        &       100         &  m \\
Diameter          & $d$     &     30                               &       20         &  mm \\
Finest pitch      & $p_{\mathrm{min}}$ &   1.7                           &       1.9        & $\mu$m \\
Profile height   &     $t$     &     8.2                                &         8.2         & $\mu$m \\
\hline
\end{tabular}
\label{table:params}
\end{table}


We chose to make the lenses of silicon since MEMS techniques are most advanced for this material and it is acceptable for lenses working in the X-ray energy range considered here. At 6.699 keV, the thickness of Si needed to give a $2\pi$ phase change ($t_{2\pi}$) is 16.8 $\mu$m. This is the peak thickness, $t$, of the profile for an ideal PFL (Figure \ref{fig:profiles}a); for a PZP, only half this thickness is required so $t = 8.4$ $\mu$m. In fact we have adopted  $t = 8.2$ $\mu$m as by reducing the depth slightly imperfect phase matching is traded for reduced absorption and a small improvement in overall efficiency is obtained, while making fabrication marginally easier.


Fabrication of the lenses was conducted in the Detector Development Laboratory (DDL) at GSFC, a fully equipped semiconductor processing facility center with Class 10 clean-room capabilities. Phase zone plate lenses (Figure \ref{fig:profiles}c) were designed using the DW2000 software tool for mask layout. Standard photolithography and advanced Deep Reactive Ion Etching (DRIE) processes were employed with a UV mask aligner (SUSS MicroTec MA-6) and a high-rate etcher (STS).

\begin{figure}
   \begin{center}
   \begin{tabular}{c}
\includegraphics[trim = 5mm 0mm 1mm 5mm, clip, width=10cm]{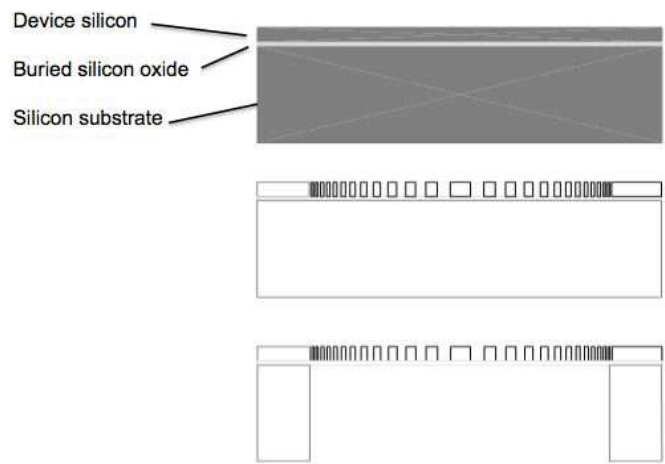}\\
   \end{tabular}
   \end{center}
   \vspace{-5mm}
   \caption[example]
   { Cross-sections of SOI wafers for illustration of the lens fabrication process. Top: SOI wafer showing the top device silicon layer, a thin insulating layer of silicon oxide, and the bottom thicker layer called the handle silicon layer;  middle: lens etched out from the device silicon layer;  bottom: support structure etched from the silicon substrate or handle silicon layer. }
  \label{fig:fab}
  \end{figure}

Each lens was fabricated from a 4-inch diameter Silicon-On-Insulator (SOI) wafer. As shown in Figure \ref{fig:fab}, an SOI wafer consists of three layers - (1) a thin Si layer on top called the device silicon layer, (2) a thin silicon dioxide (SiO$_2$) insulating layer, and (3) a thicker Si layer called the handle silicon layer.

The sequence of operations is illustrated in Figure \ref{fig:fab}. The lens pattern is first formed on the front (device) side of the SOI wafer by UV exposure of a photo-resist layer through a chromium-on-glass mask. This is followed by development and DRIE etching down to the silicon oxide insulating layer. The SOI wafer is then attached to a glass wafer with wax to protect the front-side silicon lens features. The procedure is repeated on the backside to produce the spider-web support structure, again with DRIE etching as far as the oxide layer.

This DRIE process allows slits with the required high-aspect ratios of up to 20:1 to be etched in silicon. However, the parameters of the etch process had to be tuned in order to obtain vertical silicon walls with the required width of remaining silicon between adjacent slits.  An anti-reflection coating was used to make good optical contact between the SOI wafer and the glass mask,  allowing a precise lens pattern with features at the submicron level to be obtained.

Techniques exist that allow the fabrication of lenses with multiple levels  (Figure \ref{fig:profiles}b) or even good approximations to the ideal continuous profile (Figure \ref{fig:profiles}a) -- see for example \cite{difabrizio99} and \cite{krizmanic05a}.
If lenses could be made with the same diameter and focal length as our two-level lenses but with, say,  a  4-level stepped profile, then the effective area could, in principle, be more than doubled (see Table \ref{table:eff}). Fabricating such structures would require etching features finer by at least a factor of two, and somewhat deeper.   In conjunction with the Army Research Laboratory  (ARL) in Adelphi, MD, we are investigating the feasibility of using e-beam lithography to obtain four-level profiles. Because of the long writing times involved, initially the objective is to make a sample 1 mm wide strip across the radius of a 30 mm diameter lens.

\subsection { Test Results}
\label{section:TestResults}

\begin{figure}
    \centerline{\includegraphics[width=1.0\textwidth,clip=, bb = 0 150 841 460]{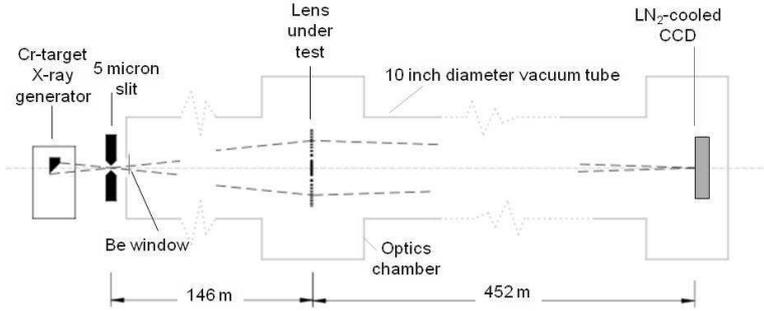} }
              \caption{Test setup at Goddard's 600-m X-ray Interferometry Testbed. A CCD X-ray camera with 13 $\mu$m pixels was used to record the magnified image of a 5-$\mu$m slit, whose width at a distance of 146 m corresponds to 7 milli-arcsec. Not to scale.
                      }
   \label{fig:TestSetup}
\end{figure}

Several lenses have been tested at the NASA-GSFC X-ray Interferometry Test\-bed. This facility provides a 600-m evacuated path between source and detector stations with an intermediate station for the focusing optics. Because of the relatively low melting point of iron and the need for high surface brightness, it was not possible to use an iron target X-ray tube, which would have produced the Fe K$\alpha$ line at 6.4 keV, close to the solar Fe {\sc xxv}  $w$ line at 6.699 keV. Instead, a chromium target X-ray tube was used, producing the Cr K$\alpha$ line at 5.411 keV.  Consequently, although the lenses were designed to have the finest pitch $p_{\mathrm{min}}$,  thickness $t$, and focal length  $f$ similar to those needed for a 20 mm diameter flight lens working at 6.699 keV, the diameter was actually 30 mm.  The focal length was chosen to be 110.4 m at the Cr line energy so that, when the lens was positioned at the `optics' station of the testbed 146 m from a 5-$\mu$m wide tungsten slit placed directly in front of the source, an enlarged image of the slit was formed in the plane of the detector, 452 m from the lens (see Figure\ \ref{fig:TestSetup}). This magnifying configuration has the advantage that the 13 $\mu$m pixels of the Roper CCD camera used as a detector were capable of resolving the image of the slit.  The camera was used in a photon-counting mode, accepting events in a narrow energy range (5.3$-$5.6 keV) containing the Cr K$\alpha$ line while excluding the K$\beta$ line at 5.947 keV.

\begin{figure}
   \begin{center}
   \begin{tabular}{c}
    \includegraphics[angle=-90, width=0.5\textwidth,clip=, bb = 200 150 500 400]{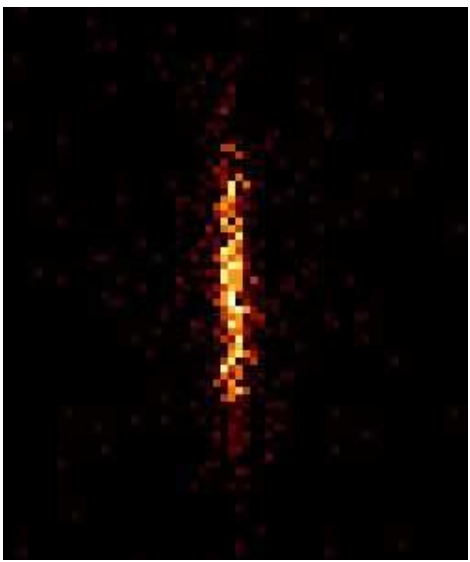}\\
   \end{tabular}
   \end{center}
   \vspace{0mm}
   \caption[example]
   { An image of a 5 $\mu$m slit obtained with the test setup in Figure \ref{fig:TestSetup}. The pixels correspond to an angular size of  6 milli-arcsec. }
   \label{fig:image}
   \end{figure}

\begin{figure}
   \begin{center}
   \begin{tabular}{c}
 \includegraphics[angle=-90, trim = 0mm 0mm 0mm 00mm, clip, width=9cm]{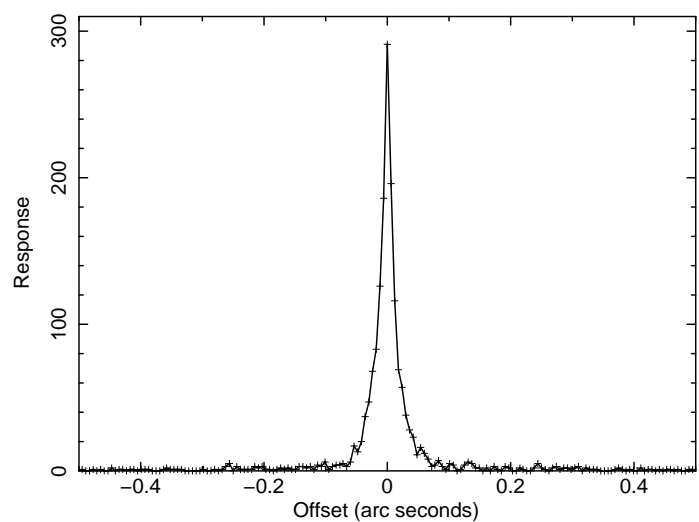}\\
   \end{tabular}
   \end{center}

   \caption[example]
   {Intensity as a function of angular offset transverse to the slit for the image in Figure \ref{fig:image}. }
   \label{fig:xsect}
   \end{figure}

 An image produced with this configuration is shown in Figure \ref{fig:image} and a cross-section though it in Figure \ref{fig:xsect}. The core of the response is 66 $\mu$m FWHM which corresponds to 30 milli-arcsec, or 28 milli-arcsec when allowance is made for broadening by the detector pixelization and the finite width (5 $\mu$m) of the slit. There is  a significant amount of power in the wings that are rather broader than the core but even measuring the width at one tenth maximum, the width of 70 milli-arcsec is better than the target of  100 milli-arcsec.

We have estimated the effective area of the lens by comparing the count rate in the peak with that when the lens was replaced by a plate with a small hole of known size. To avoid pulse pile-up effects and to allow any variation of efficiency across the lens to be investigated, some of the measurements were made with a 5 mm aperture in front of the lens. The position of the aperture could be controlled in two axes with stepper-motor drives. We find an average efficiency of 24\% with very little radial variation, leading to an effective area of  just over 2 cm$^2$. This can be compared with the theoretical value for an ideal phase zone plate of 40.5\%, which is expected to be reduced to 26.8\% when absorption, obscuration by the radial support ribs, and the fact that the profile is optimized for an energy different from the test energy, are taken into account.  A similar calculation leads to an ideal efficiency of 33.7\% at $\approx$6.7 keV.

\section{Space Mission Concept}
     \label{Sect:MissionConcept}

The major problem associated with flying such phase zone plate lenses in space and obtaining X-ray images of the Sun with sub-arcsecond resolution is the required focal length. The lenses we have developed and demonstrated have a focal length $f$ of $\approx$ 100 m.  Extensible booms up to 60 m in length have been flown in space \citep{farr07} and longer ones proposed \citep[\textit{e.g.}]{solarsail} but for such distances formation flying of two spacecraft with lenses on one and detectors on the other becomes an attractive possibility.

\subsection{Two-Spacecraft Formation Flying}

 For our purposes, the formation-flying concept requires control of the relative positions of the two spacecraft and of their orientations. Table~\ref{table:requirements} summarizes the technical requirements. Note that the spacecraft station-keeping and attitude control requirements are relatively lax. This is because images can be built up from a series of  `snapshots' and milli-arcsec imaging can be achieved providing only that milli-arcsec {\it knowledge} of the lens-detector alignment with respect to the Sun at the time of each snapshot is available. Aspect control of the individual spacecraft is required only to a fraction of a degree. The transverse station keeping requirement is dictated only by the need to keep the region of interest imaged on the detector.  The axial requirement is determined by the necessity of keeping the out-of-focus blurring to a negligible level.  Precise measurement of the actual alignment at all times then allows the image to be reconstructed taking into account the actual determined attitude without compromising the angular resolution.  If the detector is a CCD or other device that integrates between readouts and  does not allow accurate time-tagging, then there is a requirement that the alignment remain stable on the timescale of the integration -- perhaps one or a few second(s).

\begin{table}
\caption{ The principal parameters and station-keeping requirements for a solar flare X-ray imager using diffractive X-ray optics on a formation-flying mission.
}
\label{table:requirements}
\begin{tabular*}{120mm}{p{1.5in}p{0.6in}p{2.0in}} 
  \hline
\multicolumn{3}{l}{\textbf{Lens - detector requirements}} \\
Lens - detector separation & $\approx$100 m & Larger separations would allow larger lenses, but would reduce the field of view possible with a reasonable detector size \\
Pixel size for 0.1" resolution & 25 $\mu$m & X-ray CCD or X-ray pixel detector \\
Field of View & 3 arcmin & Assumes 10 cm detector diameter to cover a typical active region - could be larger. \\
\multicolumn{3}{l}{\textbf{Station keeping requirements}} \\
Control - transverse & 2 cm & Requirement is to keep image on detector - could move detector or lens independently. \\
Control - axial & 25 cm & Equivalent to 20 eV energy shift \\
Stability - transverse & $<$0.1 arcsec  \hspace{5mm}  = 25 $\mu$m at 100 m   over 1 s     & Alignment should not change between read-outs \\
Knowledge - transverse & $<$0.1 arcsec  \hspace{5mm}  = 25 $\mu$m at 100 m  & Alignment of line joining lens and detector must be known relative to the direction to Sun center to an accuracy better than the desired resolution.\\
Knowledge - axial & 10 cm & To determine the energy of the focussed X-rays to $<$10 eV. \\
\multicolumn{3}{l}{\textbf{Attitude requirement}} \\
Control and knowledge &	10 arcmin &	Tilt and tip of lens and detector are not critical \\
  \hline
\end{tabular*}
\end{table}

\subsection{Orbit Considerations}

To maintain the configuration of the two spacecraft, a quasi-continuous thrust will be needed on one, or both, spacecraft to overcome differential accelerations due to gravity gradients, radiation pressure, and drag. For spacecraft separations in the range under consideration here of order 100 m, solar gravity forces are never an important consideration. However, within the Earth-Moon environment, gravity gradient forces will generally dominate.  At a distance, D, from the center of the Earth, a spacecraft of mass $M_{\mathrm{sat}}$ will require a thrust, $\Delta F_{\mathrm{g}}$, to keep at a constant direction and separation $f$ ($f \ll D$) relative to a passive reference spacecraft. The required thrust is given by --
\begin{equation}
\Delta F_{\mathrm{g}} = 8 \left(\frac{M_{\mathrm{sat}}}{1,000\mbox{ kg}}\right) \left(\frac{D}{10,000\mbox{ km}}\right)^{-3} \left(\frac{f}{100\mbox{ m}}\right) \mbox{    mN.}
\label{eqn:diffgrav}
\end{equation}
The $D^{-3}$ term in Equation (\ref{eqn:diffgrav}) means that orbits in which the spacecraft spend most of the time far from Earth are strongly preferred. Highly eccentric orbits with periods of several days duration are probably a viable choice, though observations will be interrupted during perigee and the subsequent time necessary to reconfigure the formation. \cite{krizmanic05a} have discussed propulsion solutions for astrophysical missions with lenses having much longer focal lengths and higher thruster requirements than implied by Equation (\ref{eqn:diffgrav}).

Far from the Earth and Moon, the dominant disturbance force is likely to be differential radiation pressure. At 1 AU, this is  only 4.6 to 9.2 $\mu$N per m$^2$ of difference in effective areas of the two spacecraft normal to the Sun, depending on the reflection properties of the surfaces. Stationing the spacecraft pair close to the Sun-Earth L1 Lagrangian point would allow for almost continuous aligned observations.

We note that ESA's Proba-3 mission \citep{2010SPIE.7731E.136V, lamy10} is designed to demonstrate many of the capabilities needed for a mission to perform high angular resolution solar X-ray imaging with diffractive lenses. Proba-3 will control two spacecraft separated by 25--250 m so that a coronagraph occulter on the front spacecraft will appear directly in front of the Sun when viewed by instruments on the other spacecraft.  This technology demonstration mission will be in a 24-h eccentric Earth orbit and so the required alignment will only be maintained for limited times. The precision of the specification for the attitude determination falls only a little short of that needed to take full advantage of the capabilities of the lenses under discussion here.

A coronagraph could be a natural companion instrument to a diffractive optics X-ray imager. A lens (or multiple lenses) could be mounted within the coronagraph occulting disk. Light blocks would  ensure that they are opaque to visible/UV radiation while having negligible X-ray absorption (such light blocks will, in any case, be desirable for thermal control).

\section{Simulations of Capabilities}
     \label{Sect:sims}

\begin{figure}
   \begin{center}
   \begin{tabular}{c}
 \includegraphics[angle=-90, trim = 0mm 0mm 0mm 5mm, clip, width=6cm]{fort_22.ps}\\
 \includegraphics[angle=0, trim = 0mm 0mm 0mm 5mm, clip, width=5cm]{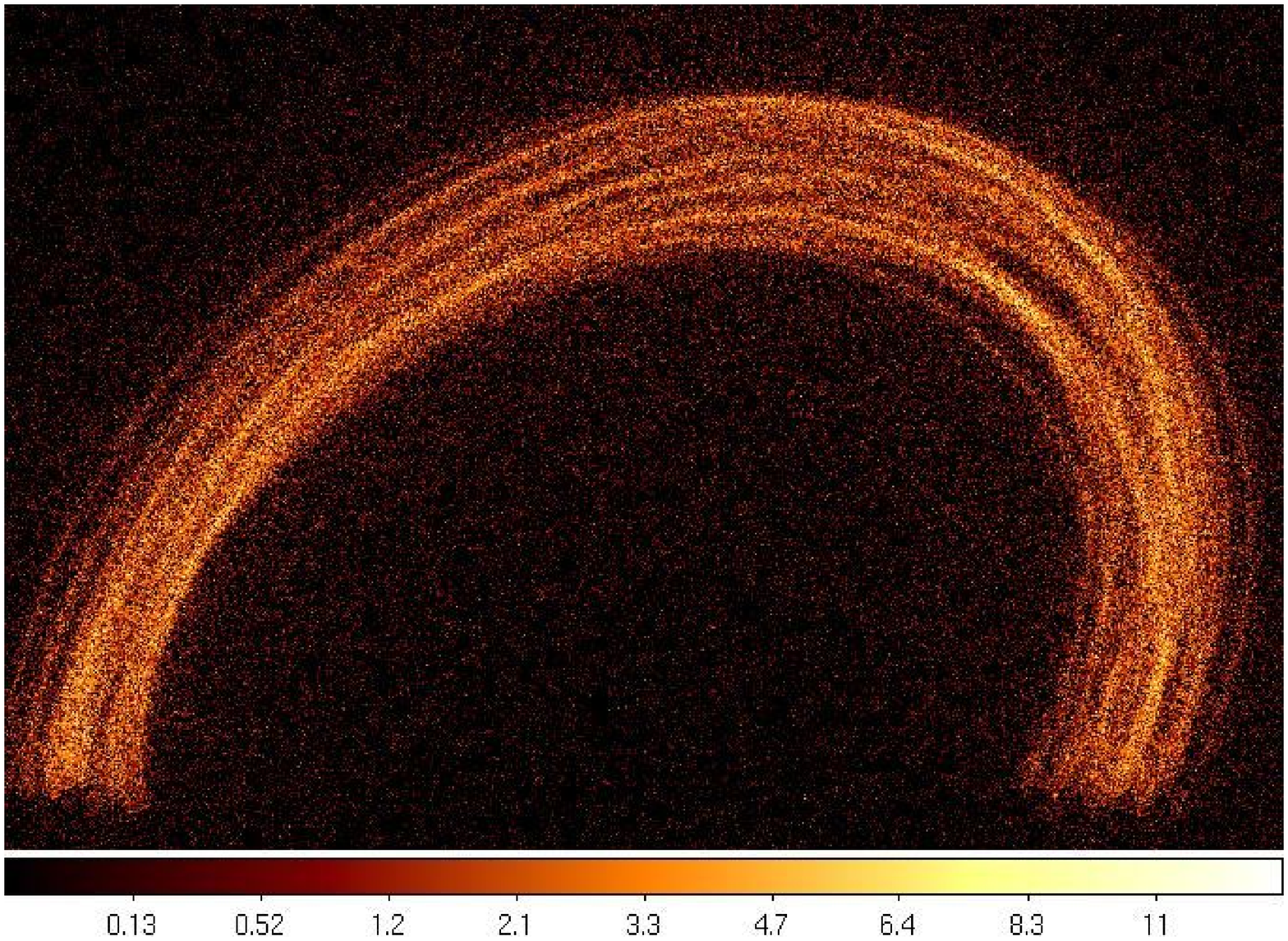}\\
 \includegraphics[angle=0, trim = 0mm 0mm 0mm 5mm, clip, width=5cm]{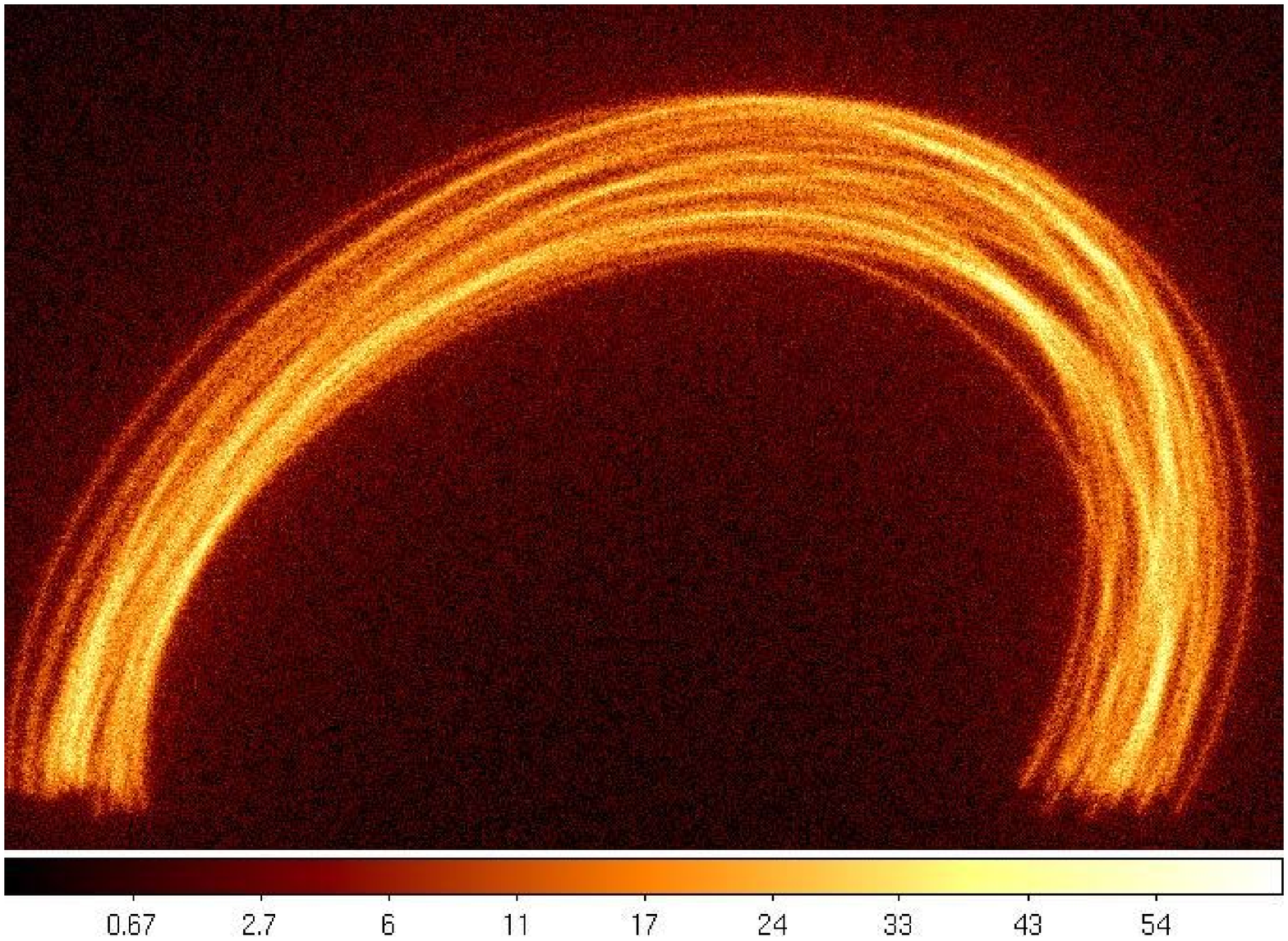}\\
   \end{tabular}
   \end{center}
   \caption[example]
   {(top) The spatial distribution model used as input to the simulations shown below. (middle) The image resulting from a 10 s (5 s effective) observation of the above source distribution.  (bottom) The corresponding image with 10 times the exposure or for a 10 times stronger flare.}
   \label{fig:SimImage}
   \end{figure}

In order to demonstrate the potential of the X-ray imaging technique considered here, we have performed Monte Carlo simulations of the operation of an instrument with a diffractive lens on one spacecraft and a CCD imaging detector on the other. We used the lens parameters for the possible flight design given in Table \ref{table:params}. The assumed lens performance is based on the lenses that we have developed and tested as described in Sections \ref{sect:fab} and \ref{section:TestResults}. We have modeled a flare of similar intensity and temperature to the 1980 June 29 event (see Figure \ref{fig:BCS_spectra}), a class M4 flare, and used the spatial distribution shown in Figure \ref{fig:SimImage}(top) in which a loop is assumed to be composed of a large number (30) of fine intertwined threads, inspired by a model illustrated in Figure 1.18 of \cite{2005psci.book.....A}.

For the spectrum of the emission, we have used a CHIANTI simulation of a plasma with coronal abundances, a temperature of 18.8 MK, and a
volume emission measure of $10^{48.84}~\mathrm{cm}^{-3}$
based on the GOES X-ray data of the flare. This  gives approximately the spectral form and normalization measured with BCS shown in Figure \ref{fig:BCS_spectra}.  In order to follow changes in the flare loop structures, we would like to be able to obtain an image in, say,  10~s. We assume that, because of the limitations of a CCD detector, a 10~s accumulation is composed of 1~s observation periods each followed by a 1~s readout time, leading to 50\% live time.

The simulations demonstrate an important consideration when making observations of relatively bright, potentially fast-changing objects. The  event rate in all the bright regions of the image precludes operation in the photon counting mode since there will be multiple photons per pixel in each 1-s frame. However, because the emission is concentrated in a narrow energy band, the total charge collected provides a good measure of the signal. For operation in this analog mode, it is important to attenuate the lower energy photons as much as possible, and  for this purpose we assume that a 25-$\mu$m-thick layer of Si and a 10-$\mu$m layer of Fe is placed in the line of sight, perhaps forming a lens substrate.

Figure \ref{fig:SimImage}(middle) shows a simulated image derived from the model image shown in Figure \ref{fig:SimImage}(top). The image quality is more affected by photon statistics than by the optics. Figure \ref{fig:SimImage}(bottom) shows a corresponding simulation of a 10 times longer observation or a 10 times stronger flare.


\section{Summary and Conclusions}
     \label{Sect:conclusions}

The results presented here show that diffractive X-ray lenses can be made that are capable of solar imaging in the Fe {\sc xxv} $w$ emission line and nearby lines. With such lenses, the capability exists for making X-ray images over narrow energy ranges with angular resolutions as fine as 0.1 arcsec. This high angular resolution and the effective area of the demonstration lenses already fabricated would allow hot plasma generated during solar flares to be studied on unprecedentedly fine spatial scales and determine the multi-threaded nature of hot coronal magnetic loops. The detailed structure of the coronal energy release sites could also be explored in detail during solar flares.

We note that the efficiency of a PZP lens can be improved by using a multilevel approximation to the ideal Phase Fresnel Lens profile rather than just the two levels of the lenses we have fabricated. In developments not reported here, we have demonstrated the capability of using grey-scale technology to generate a 4-level profile which should double the efficiency of the lens and give twice the sensitivity. Alternatively, the higher efficiency could be used to achieve a higher bandwidth with the same sensitivity by fabricating a smaller lens with a higher focal ratio. The technique proposed here can be extended to higher energies, though even longer focal distances would be needed to achieve a useful effective area unless lenses with even finer slits and higher aspect ratios can be developed. On the other hand, the lower absorption losses at higher energies may allow the bandwidth to be increased with achromatic combinations \citep{2010XROI.2010E..16S}.

Detailed design of a mounting and support structure and demonstration that the lens can withstand launch and operational environments are planned, as are further studies of the formation-flying aspects of a possible mission. More detailed development of the complete instrument concept is underway. These activities are designed to bring a high resolution X-ray imager concept to the point where it could be proposed for flight, perhaps as a science demonstrator on a formation-flying technology mission.

\begin{acks}
 We thank John Krizmanic and Keith Gendreau for supporting the lens design and testing effort, Kenneth Phillips for providing the spectra shown in Figure \ref{fig:BCS_spectra} and for help with the solar objectives, and Amil Patel, Gang Hu, and Thitima Suwannasiri for their work fabricating the lenses in Goddard's Detector Development Lab. This project was supported with funding from the Goddard Internal Research and Development (IRAD) program. CHIANTI is a collaborative project involving George Mason University, the University of Michigan (USA), and the University of Cambridge (UK).
\end{acks}

\bibliographystyle{spr-mp-sola}	
\bibliography{dxo}

\end{article}

\end{document}